\documentclass[cameraready]{Interspeech}
\usepackage{multirow}

\title{How Language-Independent Are Emotional Attributes?\\A Study on Training Data Scaling and Cross-Lingual Generalization}

\author[affiliation={1}, orcid=0009-0004-7242-8695]{Dániel}{Halmai}
\author[affiliation={1,2}, orcid=0000-0002-2864-6466]{Gábor}{Gosztolya} 

\address{
    $^1$ Institute of Informatics, University of Szeged, Szeged, Hungary\\
    $^2$ HUN-REN--SZTE Research Group on Artificial Intelligence, Szeged, Hungary
}

\email{halmaid@inf.u-szeged.hu}


\usepackage{comment}
\usepackage{subcaption}

\begin{document}

\maketitle

\begin{abstract}
Current state-of-the-art Speech Emotion Recognition (SER) systems rely on lots of training data, which is available only for a few languages such as English. For low-resource languages, using a cross-lingual model or adapting it to the target language might be an alternative to training a mono-lingual model from scratch. However, it is not clear what constitutes a `low-resource', i.e. where model adaptation loses its edge over the mono-lingual approach. We investigate this topic by training and adapting several SER models on English and Mandarin data, focusing on the emotional attributes of arousal, valence and dominance. Our results show that, for arousal, model adaptation is beneficial even with two hours of data, while for valence 10-20 hours of adaptation material is required to match the performance of a mono-lingual model trained with 100 hours of data. This indicates that task-specific model adaptation may be an effective approach in low-resource speech emotion recognition problems.
\end{abstract}

\section{Introduction}

Speech is not simply a means of communication: our emotions are constantly present in it, whether consciously or implicitly. The use of words, the pace of speech, the tone of voice, and other paralinguistic cues reveal the inner state of the speaker, allowing to automatically deduce the emotions from the speech signal (i.e. Speech Emotion Recognition, SER~\cite{schuller2018speech,wani2021comprehensive}). An automatic emotion recognition system has many potential applications, such as in education~\cite{beccaro2024analysisoforalexams}, in the pathological field~\cite{pepa2023automaticemotionrecognition} and in general human-machine interaction~\cite{spezialetti2020emotionrecognition,okayama2025datasetconstruction}. This field has been intensively studied for well over a decade~\cite{schuller2011recognising}, while in the recent years its development was accelerated by the use of deep neural networks and self-supervised learning~\cite{wagner2023dawnofthetransformerera}.

An obvious limit 
in automatic speech emotion recognition is the amount of training data available. Although in the past few years large corpora have became available (most notably, MSP-Podcast~\cite{MSP_Podcast}), this is not the case with most scenarios, e.g. for most languages~\cite{emodb,sztaho2011automaticclassification}. To build a SER system with limited resources, there are three straightforward options. First, one can train a model on the data available in the target language. Second, one can rely on a model trained on some large corpus, even if it is of a different language, and hope that this cross-lingual (or even cross-cultural) setup will not hinder the performance too much on the target language. The third option is to use both kinds of data combined in some way -- for example by taking the model trained on the `resourceful' language and further fine-tuning it on the target language (i.e. domain adaptation~\cite{chen2015deepsemisupervised}). Generally, the third option can be expected to lead to the best performance, but it is unclear how much training data will be required in the target language to effectively adapt the original model. Furthermore, the different models might deliver uneven performance depending on the emotional content of the actual speech utterance; for instance, the cross-lingual performance might differ markedly for different emotional categories or attributes (e.g. for arousal and valence) 
since emotion expressions may differ from one language to another~\cite{lim2016cultural}.

To answer these questions, in this study we trained emotion recognition models on two quite different languages: English and Taiwanese Mandarin. We measured the performance of the trained models on the Mandarin corpus in each case, testing several training setups. As straightforward solutions, we trained models on only one language, using roughly 100 hours of data in both cases (mirroring cross-lingual and intra-lingual, but `resourceful' scenarios). To simulate low-resource settings, we selected several smaller subsets of the Mandarin dataset, and trained SER models both from scratch and by fine-tuning the networks originally trained on all 100 hours of the English data. We focused on the emotion dimensions; instead of summarizing model performance by one single metric value, we examined the arousal, valence and dominance attributes one by one.

\section{Related Work}

Most SER studies, even if investigate multiple languages, do not present cross-corpus experiments, just report the performance of an approach on several languages~\cite{meftah2016emotional,singh2023decoding,porjazovski2024improvedspokenemotion,hu2025research}. One of the reasons for this might be that 
it is hard to convert the emotional categories into a shared set of labels. A few studies performed that, though; for example, Ma et al. kept only four categories (angry, happy, neutral and sad) and measured the intra-corpus and cross-corpus performance of several architectures on 32 datasets in 14 languages~\cite{ma2024emobox}. Heracleous and Yoneyama went even further by training a joint multi-lingual model for Japanese, English and German~\cite{heracleous2019comprehensive}. 
Pastor et al. trained HuBERT and WavLM models for the same four categorical emotions on IEMOCAP extended with a few minutes from EmoDB and RAVDESS. They found that including more in-corpus training data might drastically increase the performance~\cite{pastor2023cross}.

If one investigates emotional attributes (such as arousal and valence) instead of categorical class labels, performing cross-corpus (and thus cross-lingual) experiments becomes easier. (Although, to avoid scaling issues, it is standard to convert the attributes to a binary set of low and high (or to the triplet of low, neutral and high).) Eyben et al., back in 2010, evaluated several subsets of a large hand-crafted attribute set on three German and one English datasets both in an intra-corpus and in a  cross-corpus manner~\cite{eyben2010crosscorpus}. Later, Schuller et al. repeated a similar set of experiments on six datasets, using both categorical labels and arousal / valence (converted to binary)~\cite{schuller2015crosscorpusacoustic}. A similar study by Feraru et al. evaluated the cross-lingual performance of models in terms of (binary) arousal and valence prediction for eight languages, investigating linguistic groups as well~\cite{feraru2015cross}.

The study which is the closest to our research question is that of Neumann et al.~\cite{neumann2018cross}. They used the IEMOCAP (English) and Recola (French) corpora, focusing on the attributes of arousal and valence (again, converted to binary). Besides training mono-lingual and multi-lingual models, they tested model adaptation, i.e. fine-tuning a model on the target language. Their results are mixed, though: regarding arousal, the performance of the adapted model roughly matched that of the multi-lingual DNN, but for valence most results were around the level of random guessing.

In general, all these works used small datasets (even IEMOCAP is only around 12 hours), and very few experimented with cross-lingual model adaptation. When investigating attributes like arousal and valence, they reduced the task to a binary classification one, discarding important details in model performance. Predictions of valence were of low quality, perhaps since most of the works preceded the transformer era. Finally, the few experiments involving model adaptation were either not interested in the amount of adaptation data~\cite{neumann2018cross}, or they used only a few minutes~\cite{pastor2023cross}. In contrast, here we (i) rely on a current backbone with a transformer architecture (i.e. WavLM), (ii) utilize relatively large amounts of training data and several variations of adaptation data,
and (iii) adapt the deep learning models in a cross-lingual manner (while also training them from scratch).
\begin{table}[!t]
\centering
\caption{Main properties of the MSP Podcast and BIIC Podcast corpora used in our experiments}
\label{table_dataset}
\renewcommand*{\arraystretch}{1.1}
\begin{tabular}{|l|l||r|c|c|}
\hline
\multicolumn{2}{|l||}{}                & \textbf{No. of~} & \multicolumn{2}{c|}{\textbf{Duration}}\\
\cline{3-5}
\multicolumn{2}{|l||}{\textbf{Subset}} & \textbf{chunks} & \textbf{Mean} & \textbf{Total}      \\
\hline\hline
\multicolumn{2}{|l||}{MSP train} & 65\,205 & 5.75 s & 104.19 h \\ 
\hline
\multirow{6}{*}{BIIC train}
                & ~~~~1h  & 487   & 7.57 s & ~~~~1.02 h \\ \cline{2-5}
                & ~~~~2h  & 974   & 7.56 s & ~~~~2.05 h \\ \cline{2-5}
                & ~~~~5h  & 2\,436 & 7.54 s & ~~~~5.10 h \\ \cline{2-5}
                & ~~10h & 4\,873 & 7.55 s & ~~10.21 h \\ \cline{2-5}
                & ~~20h & 9\,747 & 7.54 s & ~~20.41 h \\ \cline{2-5}

                & 100h & 48\,737 & 7.56 s & 102.38 h \\ \hline

\multicolumn{2}{|l||}{BIIC dev} & 10\,825 & 7.41 s & ~~22.29 h \\ \hline
\multicolumn{2}{|l||}{BIIC test} & 10\,312 & 7.87 s & ~~22.54 h \\ 
\hline
\end{tabular}
\end{table}


\section{Data}

In our work, we used the BIIC-Podcast corpus~\cite{BIIC_Podcast} and the MSP-Podcast corpus~\cite{MSP_Podcast}. Although these datasets differ in their language (Taiwanese Mandarin and English, respectively) and were created by different research teams, these datasets are intentionally very similar. Both datasets include naturalistic spontaneous recordings, use continuously expanding, non-static data collection, and are annotated following the same protocol.

The audio clips of both datasets are annotated with emotional attributes and with a categorical emotion. The emotional attributes are arousal, valence and dominance, each annotated on a 1-7 Likert-type scale. The categorical emotion can be anger, contempt, disgust, fear, happiness, neutral, sadness, surprise and other. (Clips where the annotators could not reach a clear majority were labeled as a tenth category: mixed.) In this paper, we focus on the emotional attributes.

\subsection{MSP-Podcast}

To the best of our knowledge, MSP-Podcast is the largest speech emotion dataset today. It consists of freely available podcast recordings in English. Each audio clip was annotated by at least five workers and the final score was determined by the average value. The authors also published individual responses. We used v1.11 of the corpus, and only those clips which had a categorical emotion label among the eight emotions (i.e. discarded utterances with the ``Other'' and ``Mixed'' labels), leaving cca. 65k recordings with a total duration of 104 hours.

\subsection{BIIC-PodCast}

The BIIC-Podcast corpus is a Taiwanese Mandarin dataset. Similar to the MSP-Podcast, the recordings were collected from publicly available online sources that provide creative commons (CC) licenses under CC-BY or CC-BY-SA; however, the authors collected not only podcasts, but also monologues, conversations and other audio materials. Each clip was annotated by 3 to 5 annotators (the final score was determined by the average value); however, the individual votes are not available in the corpus. In this corpus, the number of audio records with ``Other'' and ``Mixed'' label is negligible. It contains 70k audio recordings, of which 81\% is the training set, with 102 hours. 

Table \ref{table_dataset} presents the details of the two datasets.

\section{Methods}
\begin{figure*}[t]
\centering

\begin{subfigure}{0.32\textwidth}
    \centering
    \includegraphics[width=\linewidth]{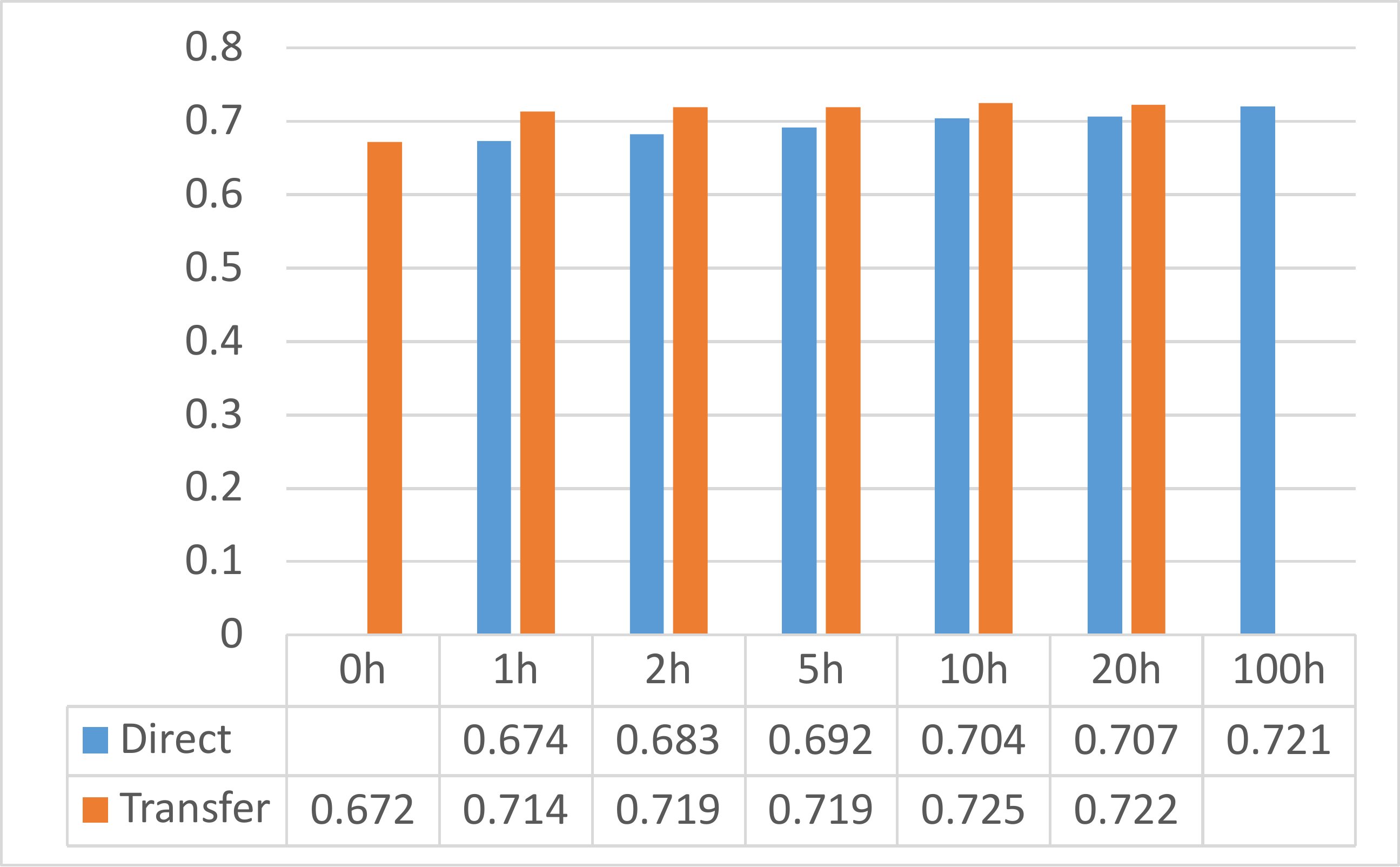}
    \caption{Pearson – Arousal}
\end{subfigure}
\hfill
\begin{subfigure}{0.32\textwidth}
    \centering
    \includegraphics[width=\linewidth]{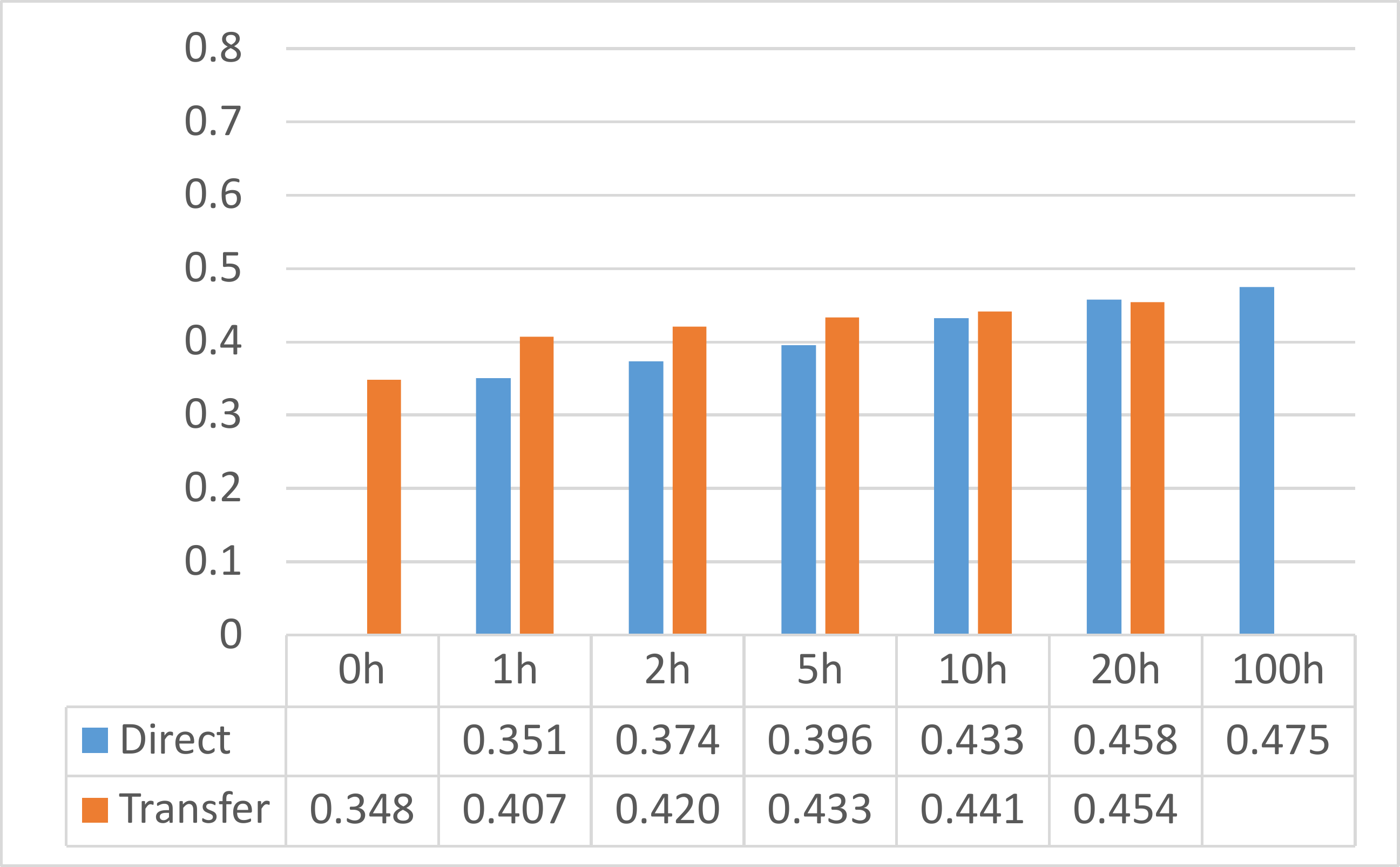}
    \caption{Pearson – Valence}
\end{subfigure}
\hfill
\begin{subfigure}{0.32\textwidth}
    \centering
    \includegraphics[width=\linewidth]{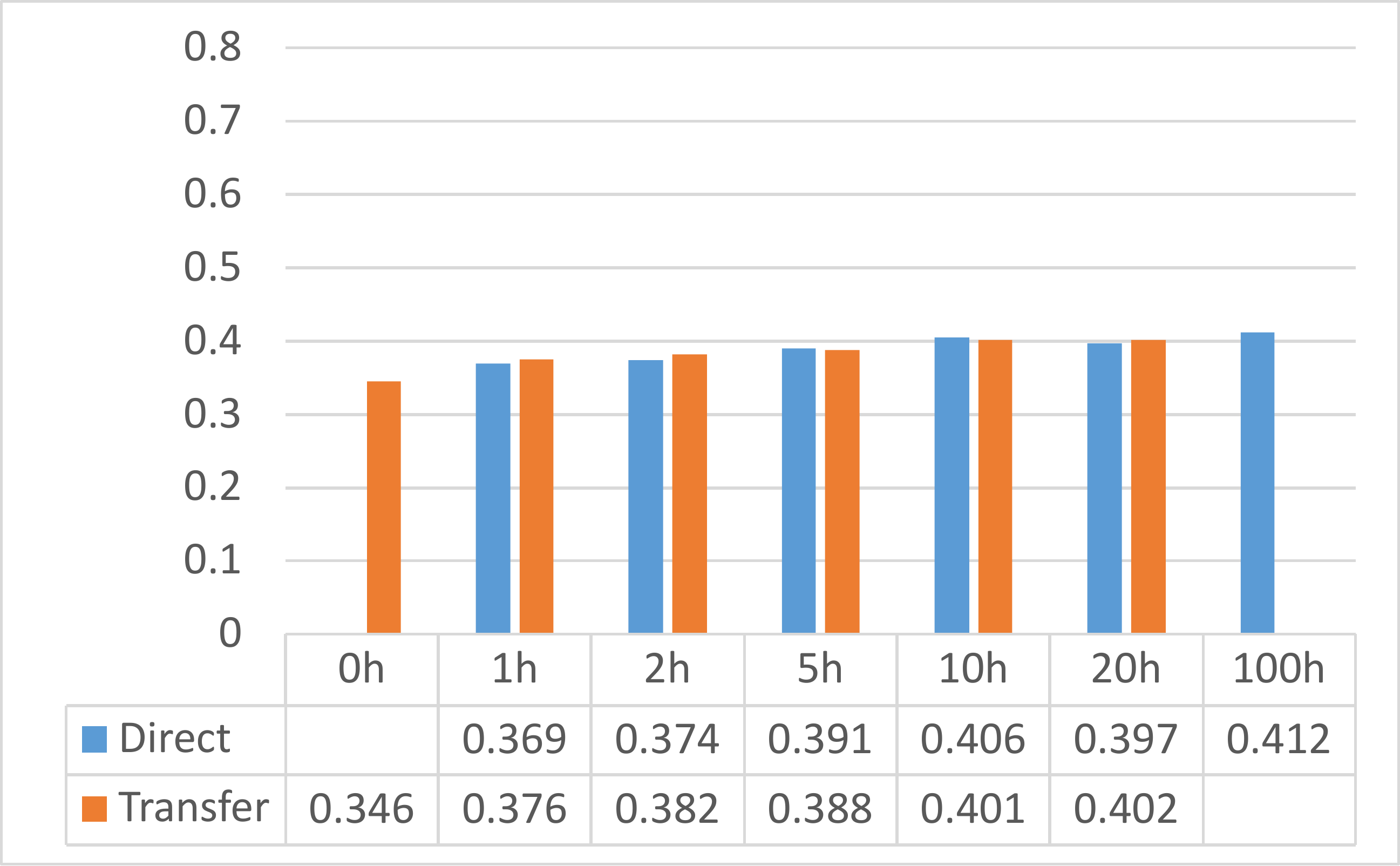}
    \caption{Pearson – Dominance}
\end{subfigure}

\vspace{0.4cm}

\begin{subfigure}{0.32\textwidth}
    \centering
    \includegraphics[width=\linewidth]{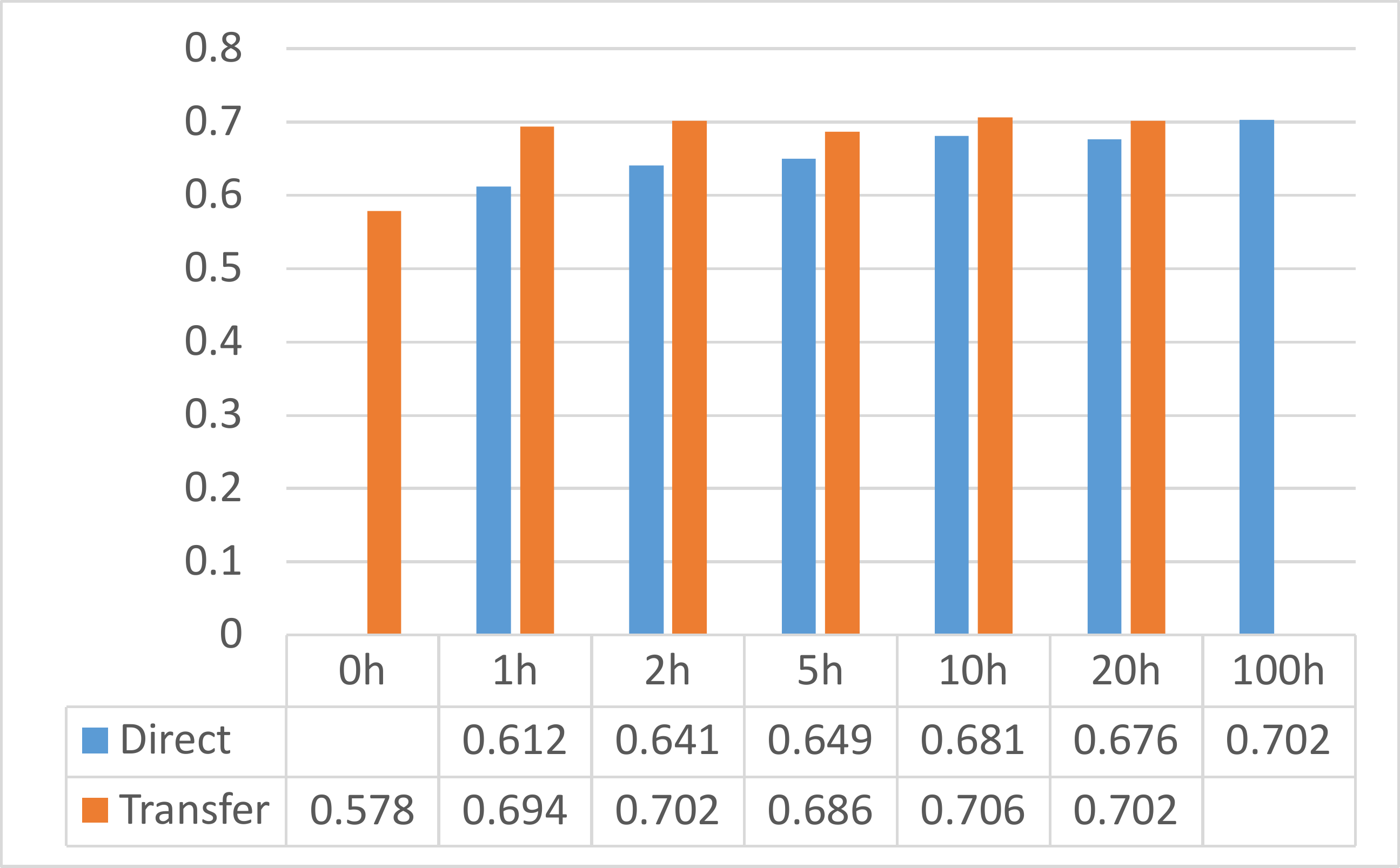}
    \caption{CCC – Arousal}
\end{subfigure}
\hfill
\begin{subfigure}{0.32\textwidth}
    \centering
    \includegraphics[width=\linewidth]{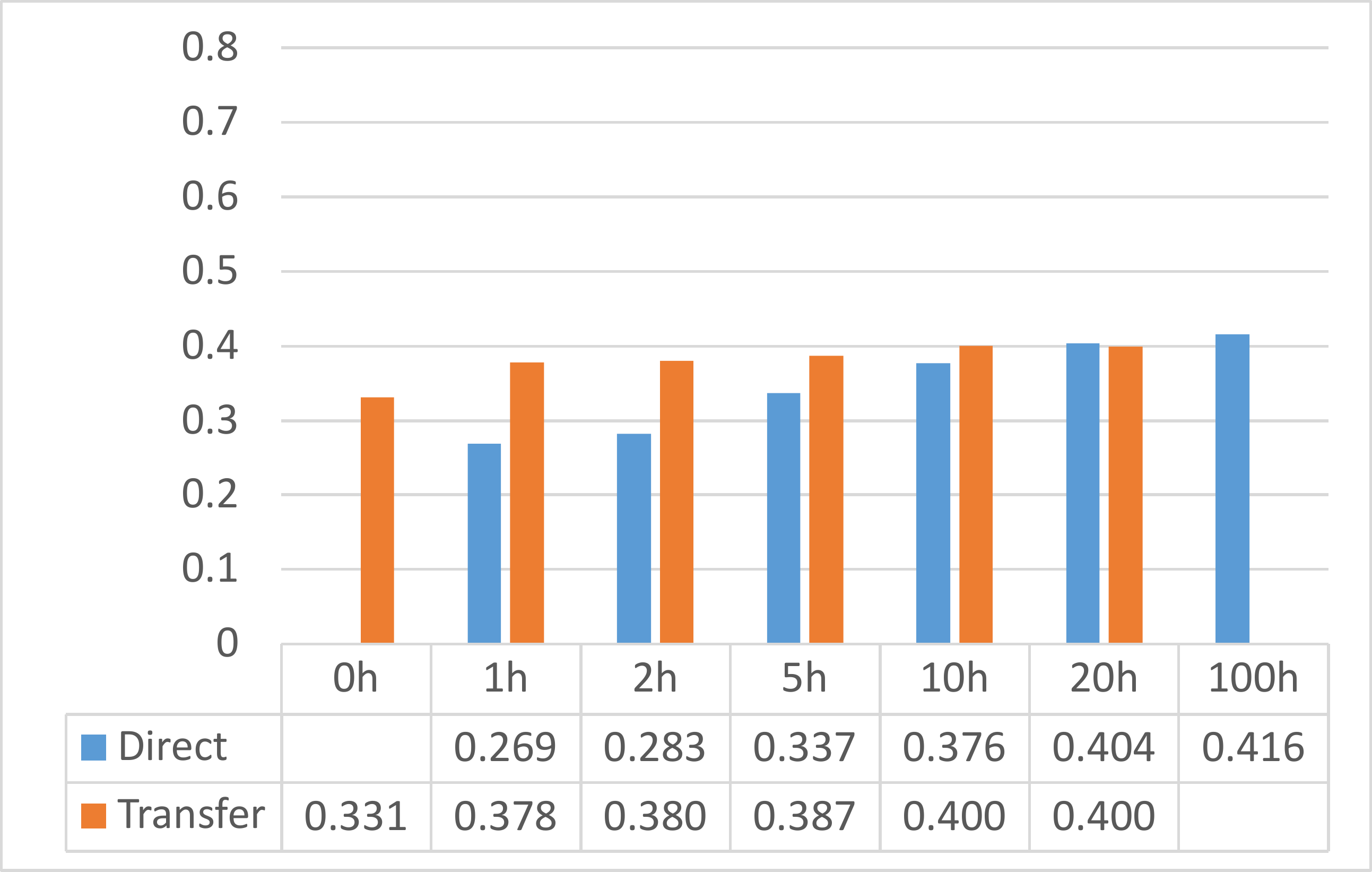}
    \caption{CCC – Valence}
\end{subfigure}
\hfill
\begin{subfigure}{0.32\textwidth}
    \centering
    \includegraphics[width=\linewidth]{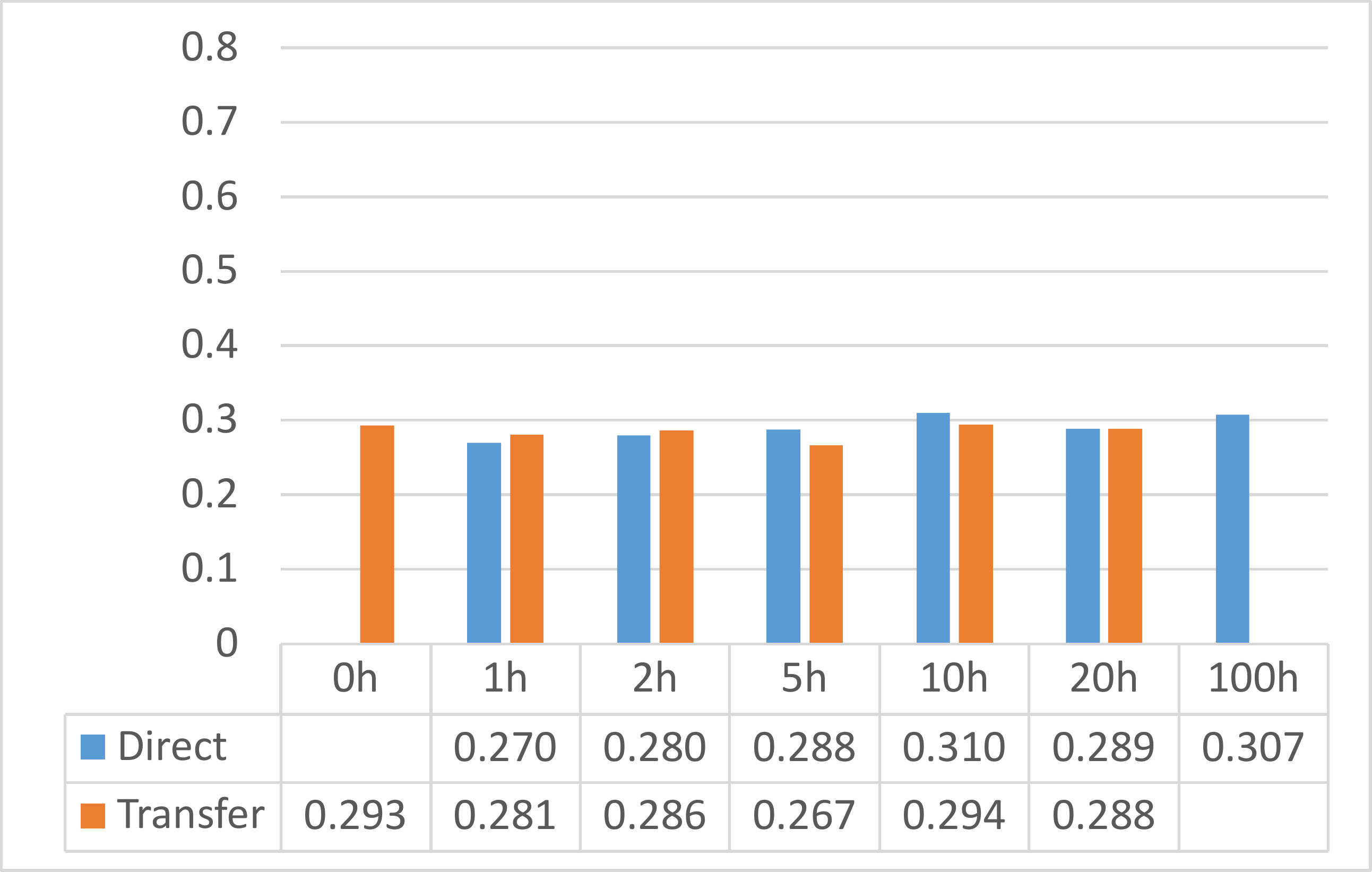}
    \caption{CCC – Dominance}
\end{subfigure}

\caption{Pearson's correlation and CCC results on the {\bf development} set.}
\label{fig:dev_results}
\end{figure*}

\subsection{WavLM}

To create our models, we chose WavLM as the backbone model. Three types of WavLM model configurations are available: WavLM Base, WavLM Base+ and WavLM Large~\cite{wavlm}. The main differences between the models are the number of parameters, the number of transformer layers and the amount of data used during pre-training. We used the WavLM large\footnote{\tt microsoft/wavlm-large} due to its performance, and because it is widespreadly used for emotion recognition~\cite{goncalves2024odyssey,zgorzynski2025multitasklearning,lertpetchpun2025developingahighperformance}. This model has 316M parameters, 24 transformer layers, and it was pre-trained by using over 94k hours of training data.

\subsection{Models}

To examine the effect of limited training data, we selected subsets of the BIIC-Podcast corpus of roughly 1, 2, 5, 10, and 20 hours, and trained separate models on each subset from scratch. In the following, these models are referred to as the `Direct' models. We also trained a model on the whole training data (102h) of the BIIC-Podcast corpus (`Direct-100h').

To explore language independence, we first trained a WavLM Large model on the MSP-Podcast corpus. This English SER model was then fine-tuned on the BIIC-Podcast subsets to assess the effect of cross-lingual adaptation; these models are referred to as 'Transfer' models. To follow the naming convention, the basic English model will be denoted as `Transfer-0h'.

\section{Experimental Setup}

To create our models, we used the SpeechBrain framework, which is an open-source, all-in-one toolkit that facilitates research and development in speech processing~\cite{speechbrain}.

The convolutional layers of the backbone model were frozen, and the network was fine-tuned for 20 epochs. The task-specific layers were trained using the Adam optimizer with a learning rate of $5\times 10^{-5}$. The model was trained on a regression task with three output neurons, corresponding to the three emotional attributes (i.e. arousal, valence and dominance). The loss function was mean squared error (MSE); finally, model checkpoints were selected based on the minimum of MSE on the development set.

To compare the results, we used Pearson correlation and Concordance Correlation Coefficient (CCC), both evaluated on each emotional attribute. To reduce the effect of randomness, we trained each model using 5 different random seeds, and we report the mean metric scores obtained. This setup also allows us to compare the significance of the differences between model performances; we used the Mann-Whitney $U$ test~\cite{mann1947onatest} (also known as the Wilcoxon rank-sum test).

\section{Results}

Fig.~\ref{fig:dev_results} shows the Pearson's correlation and the CCC results on the development set by emotional attributes. In the 0h column, we can observe the pure performance of the English model on the Taiwanese dataset, while the 100h column corresponds to the model trained only on the BIIC-Podcast corpus. Comparing these two monolingual models, the performance difference is roughly the same for the three attributes and the two metrics (around 0.05$\ldots$0.1 absolute). The exception is CCC for the dominance attribute, where the two models produce quite similar scores (0.293 vs. 0.307), probably due to the relatively low CCC value of the Mandarin-trained model compared to the Pearson's Correlation value (i.e. 0.307 vs. 0.412). It is also apparent that the values measured for valence and dominance are a lot worse than for arousal, even for the within-language model (i.e. Direct-100h), which indicates that these attributes are harder to predict for Mandarin. (This is not the case, for example, for English~\cite{busso2025msppodcast20}.)

Examining the cases with limited training data, for the {\bf arousal} attribute the Transfer models led to better results than the Direct models in all cases, and the difference is statistically significant (i.e. $p < 0.05$) for any amount of in-language training data and for both metrics. With 10 hours of Mandarin data used for model adaptation (i.e. `Transfer-10h'), we can even surpass the performance of the Direct-100h model. However, neither this nor any other Transfer model led to a statistically significant improvement ($p >$~0.05 in all cases). Yet this means that, at least on the development set, even one hour of in-language (i.e. Taiwanese Mandarin) data is enough to adapt a DNN model previously trained for English to attain the performance of a model trained with 100 hours of Mandarin speech.

In contrast, for {\bf valence} and {\bf dominance} no `Transfer' model was able to surpass the mono-lingual case with 100 hours of data (i.e. `Direct-100h'); on the contrary, all scores were significantly worse (with the exception of CCC with 2, 10 and 20 hours used for adaptation). Model adaptation kept its edge over training the models from scratch when the training data was at most 2 hours for the valence attribute, though. This underlines that arousal seems to be relatively language-independent, but the other two attributes are really specific to the given language.

\begin{figure*}[t]
\centering

\begin{subfigure}{0.32\textwidth}
    \centering
    \includegraphics[width=\linewidth]{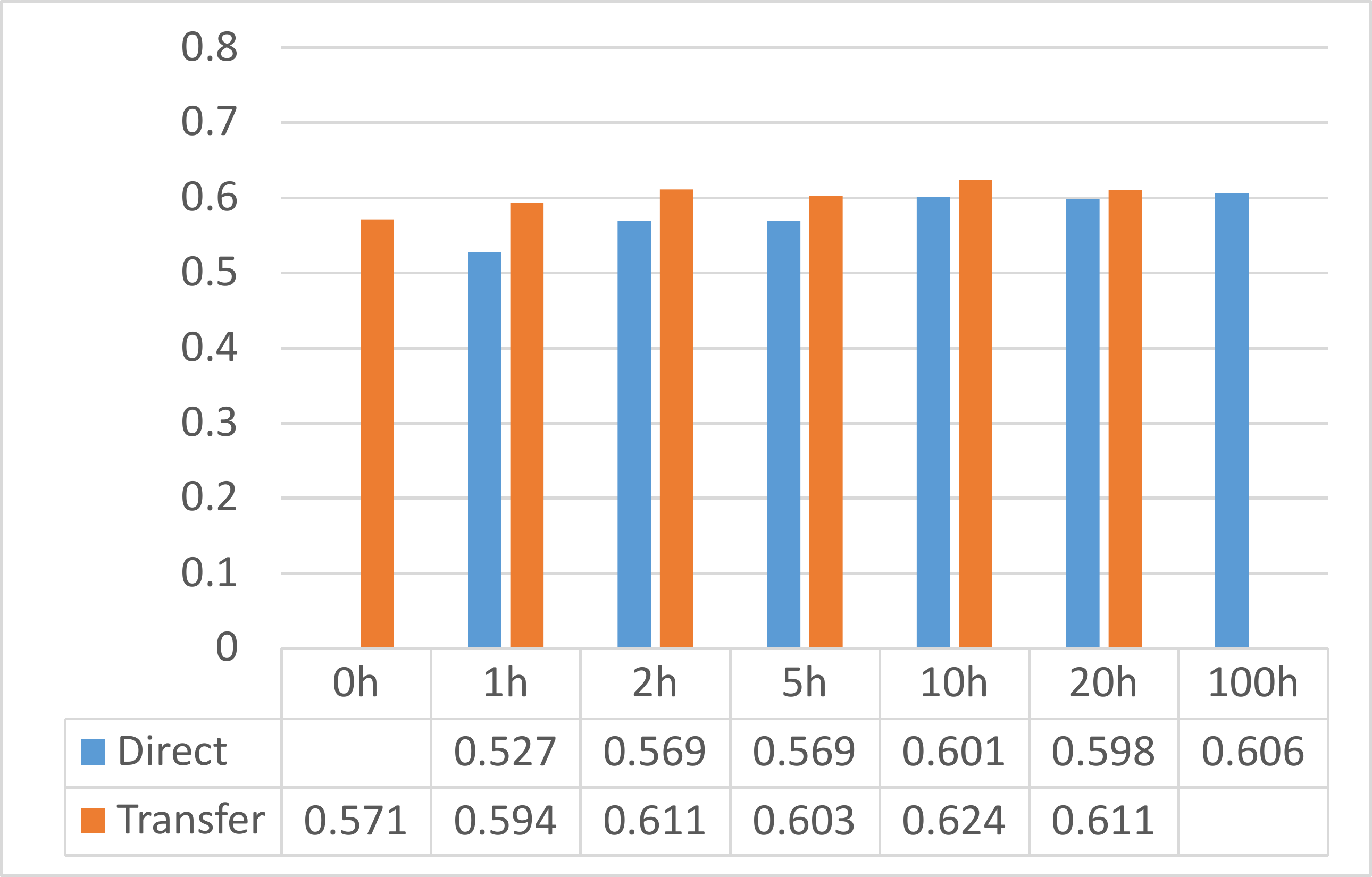}
    \caption{Pearson – Arousal}
\end{subfigure}
\hfill
\begin{subfigure}{0.32\textwidth}
    \centering
    \includegraphics[width=\linewidth]{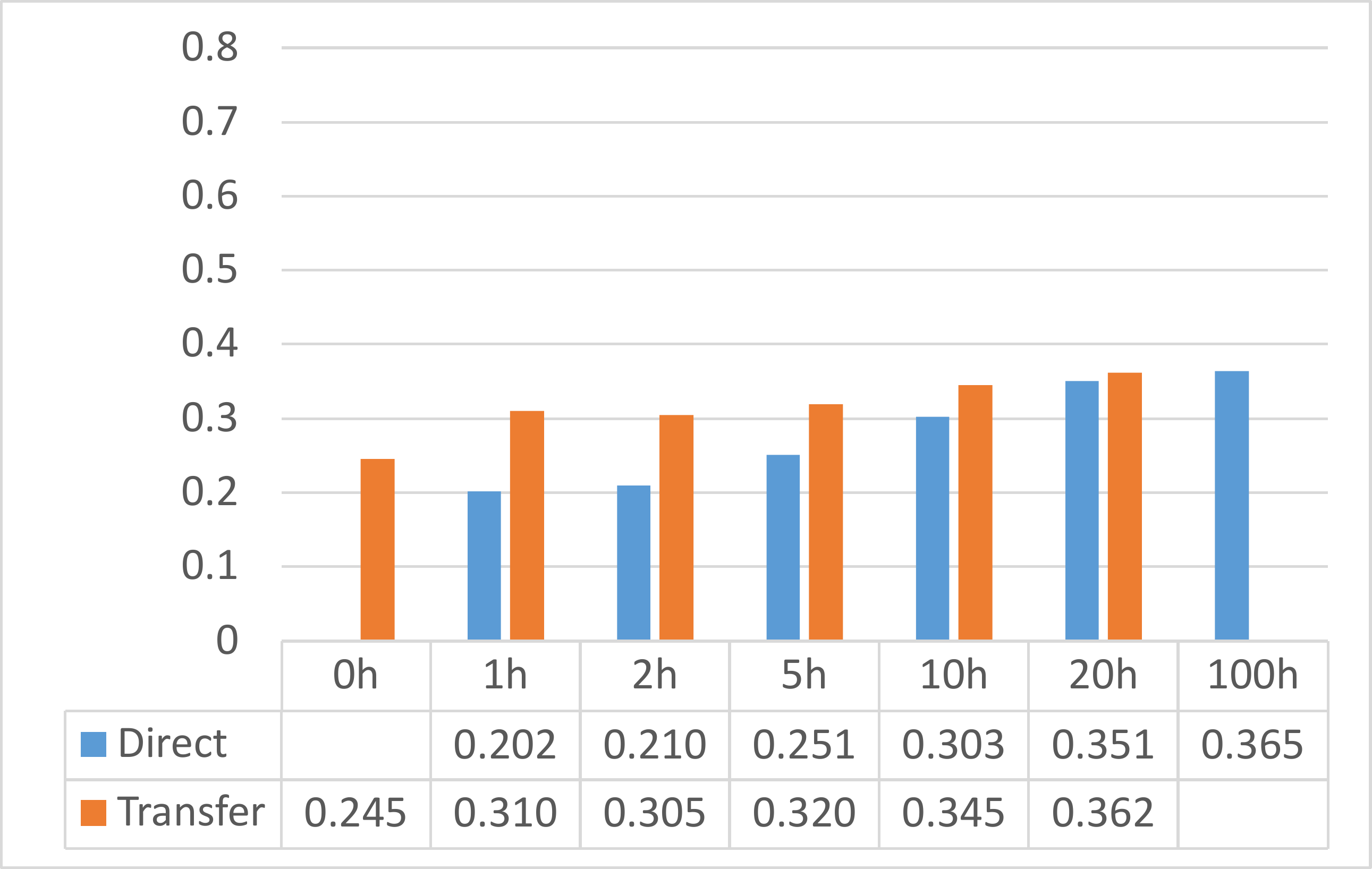}
    \caption{Pearson – Valence}
\end{subfigure}
\hfill
\begin{subfigure}{0.32\textwidth}
    \centering
    \includegraphics[width=\linewidth]{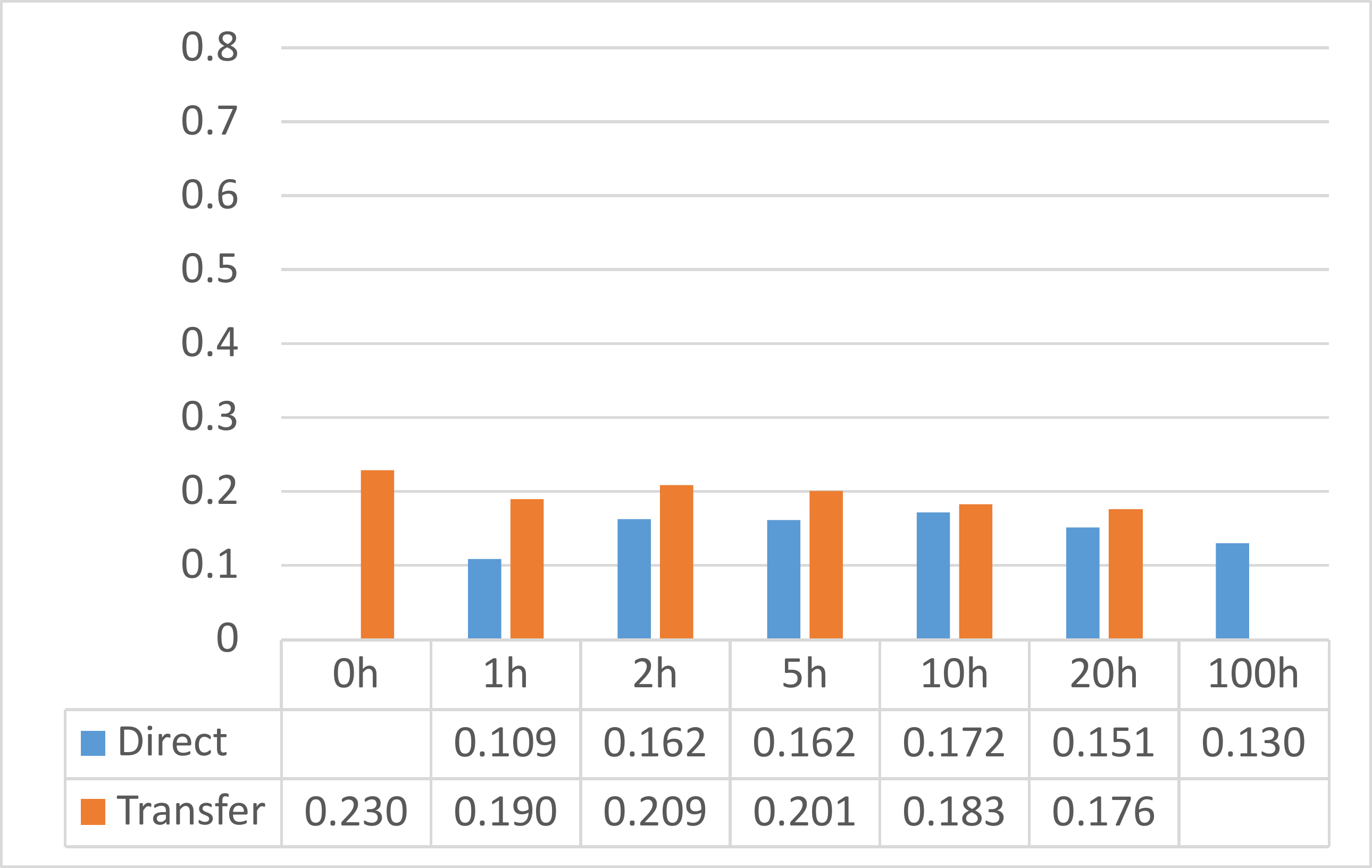}
    \caption{Pearson – Dominance}
\end{subfigure}

\vspace{0.4cm}

\begin{subfigure}{0.32\textwidth}
    \centering
    \includegraphics[width=\linewidth]{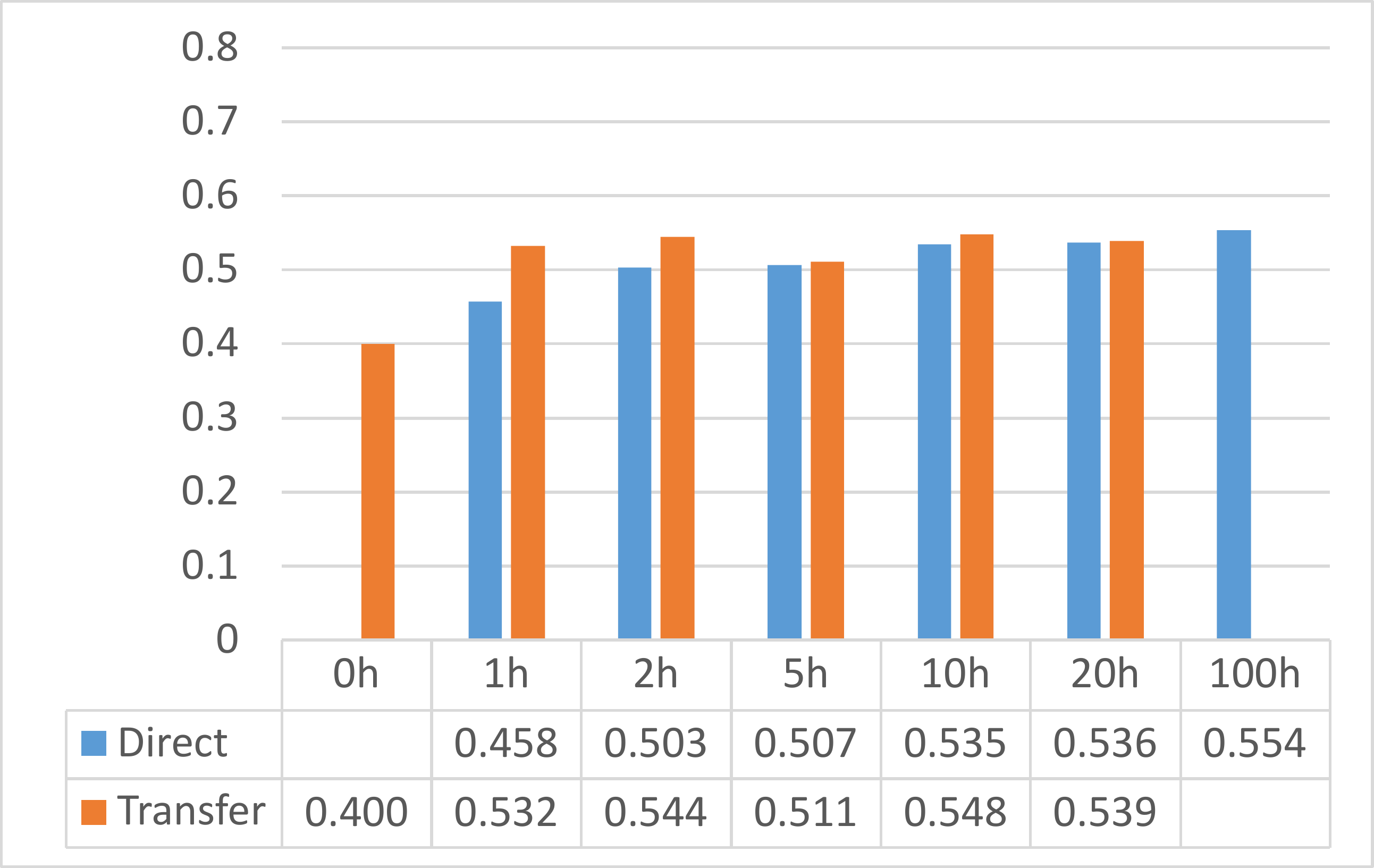}
    \caption{CCC – Arousal}
\end{subfigure}
\hfill
\begin{subfigure}{0.32\textwidth}
    \centering
    \includegraphics[width=\linewidth]{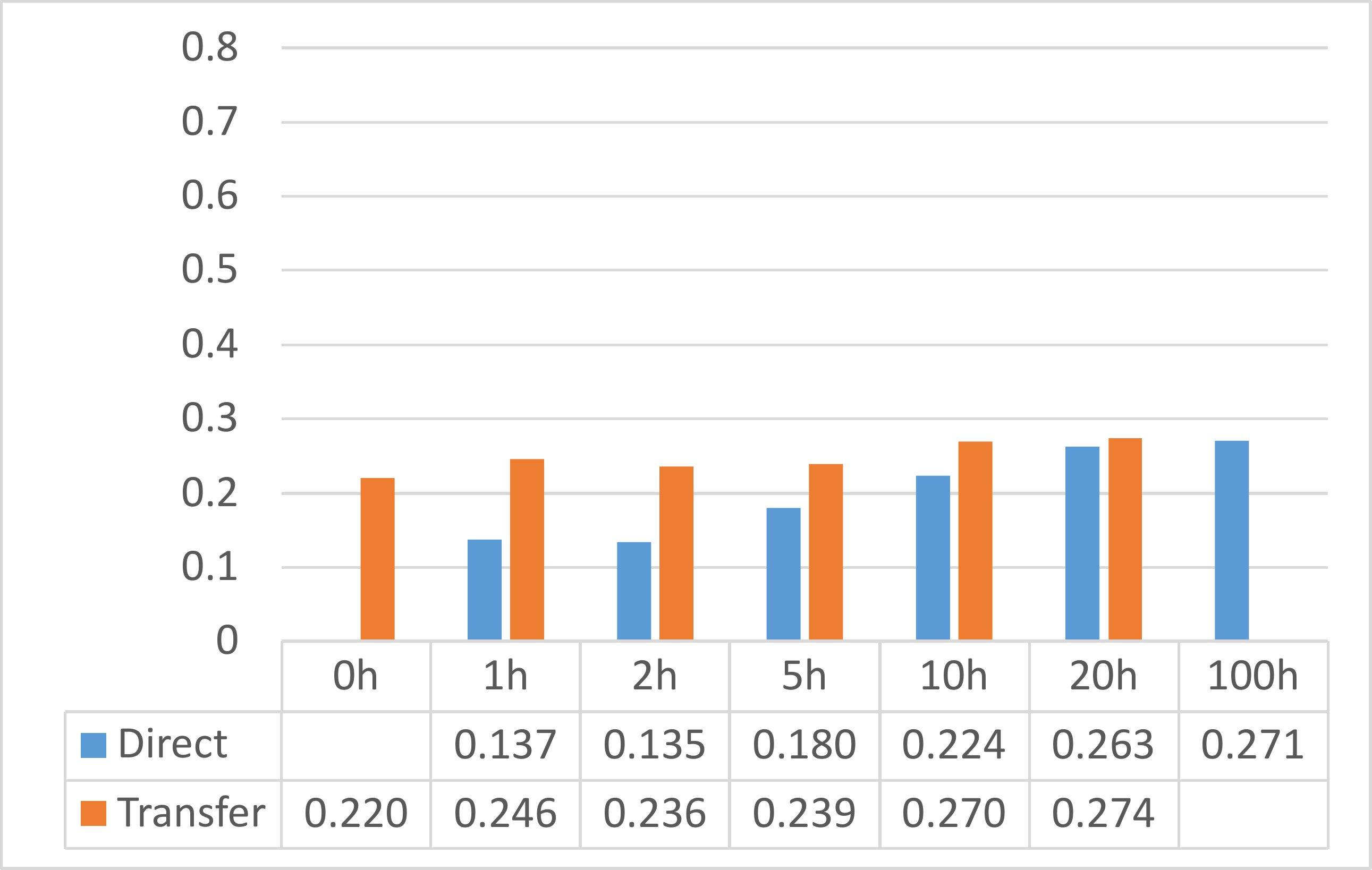}
    \caption{CCC – Valence}
\end{subfigure}
\hfill
\begin{subfigure}{0.32\textwidth}
    \centering
    \includegraphics[width=\linewidth]{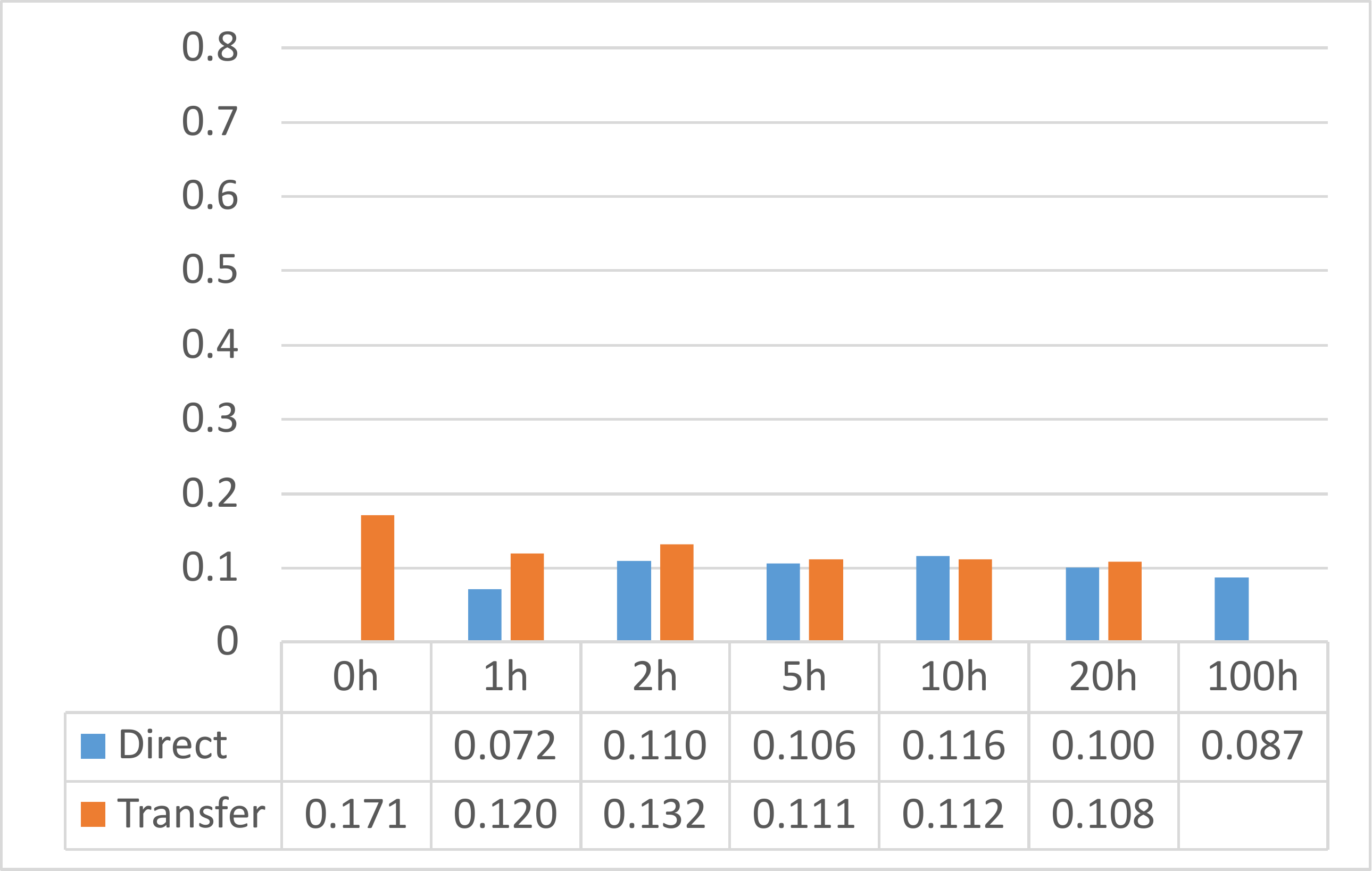}
    \caption{CCC – Dominance}
\end{subfigure}

\caption{Pearson's correlation and CCC results on the {\bf test} set.}
\label{fig:test_results}
\end{figure*}

\begin{table}[!t]
\centering
\caption{Test set performance of the models trained with the in-language data on 100 hours (`Direct-100h') and the best adapted model (`Transfer', chosen based on development set performance)}
\label{table_best_cases}
\renewcommand*{\arraystretch}{1.1}
\resizebox{\columnwidth}{!}{
\begin{tabular}{|l|l|r||c|c|}
\hline
{\bf Attribute} & {\bf Approach} & {\bf Hrs.} & {\bf Pearson} & {\bf CCC} \\
\hline\hline
\multirow{3}{*}{Arousal}
        & Direct    & 100h & 0.606 & 0.554 \\
\cline{2-5}
        & Transfer  &  10h & 0.624 & 0.548 \\
\cline{2-5}
        & \multicolumn{2}{l||}{Significance level} & $p <$ 0.01 & $p = $ 0.547 \\
\hline\hline
\multirow{3}{*}{Valence}
        & Direct    & 100h & 0.365 & 0.271 \\
\cline{2-5}
        & Transfer  &  20h & 0.362 & 0.274 \\
\cline{2-5}
        & \multicolumn{2}{l||}{Significance level} & $p = $ 0.547 & $p = $ 0.691 \\
\hline\hline
\multirow{3}{*}{Dominance}
        & Direct    & 100h & 0.130 & 0.087 \\
\cline{2-5}
        & Transfer  &  10h & 0.183 & 0.112 \\
\cline{2-5}
        & \multicolumn{2}{l||}{Significance level} & $p <$ 0.01 & $p <$ 0.01  \\
\hline
\end{tabular}
}
\end{table}
Figure~\ref{fig:test_results} presents the Pearson correlation and CCC results on the test set for the three emotional attributes. Overall, a similar  trend can be observed as on the development set; however, the absolute scores are consistently lower. Model adaptation leads to significantly better values than direct training in all six cases with only 1 and 2 hours of adaptation data ($p < 0.01$), and this holds for some other cases as well (in all cases with the arousal attribute when using Pearson's correlation, for 5 hours with dominance and Pearson's, and for 5 and 10 hours for valence with both evaluation metrics).

The superiority of model adaptation is clearly shown by the fact that the performance of the within-language model (i.e. `Direct-100h') could be achieved with significantly less training data, while sometimes it could even be surpassed. That is, the `Transfer-10h' model produced a significantly better Pearson's score for arousal, while there was no difference in the performance measured by CCC. For the valence scores, model adaptation using 20 hours was on par with the within-language model trained on 100 hours, while for dominance this held for all `Transfer' cases. The reason for the latter is probably the low Pearson's and CCC values obtained by using the Mandarin training data, for which we have no plausible explanation (except perhaps label error).

Table~\ref{table_best_cases} shows the performance of the best models (chosen based on the scores of the development set) on the test set. Clearly, by using only 10 hours of data for model adaptation, significant improvements could be achieved for the arousal and dominance attributes. The exception is the CCC metric for arousal, where the CCC values (averaged over five DNN models) were practically the same. Regarding valence, it took more Mandarin data (20 hours), but it was possible to reach the same level of performance by model adaptation as the mono-lingual Mandarin model trained on 100 hours. These findings are promising, since they suggest that effective speech emotion recognition systems can be trained on limited amounts of data by using similar training material on a different language.

\section{Conclusions and Discussion}

In this study we investigated dimensional emotion recognition in low-resource settings. To this end, we trained a WavLM Large model on 100 hours of English data, and trained it further with significantly smaller amounts of Mandarin recordings. These adapted models were compared to DNNs trained from scratch only on Mandarin speech.

We found model adaptation to be quite successful: for tiny amounts of data (i.e. 1 and 2 hours) in the target language, it led to statistically significant improvements over the mono-lingual approach (using the same Mandarin data) for all three emotion dimensions (i.e. arousal, valence and dominance), and kept its edge for specific attributes (most notably, arousal and valence) even when we used more Mandarin data. With only 2 hours of Mandarin speech, the adapted model was on par with the monolingual model trained on 100 hours in arousal and dominance prediction, while with 10 hours it could even outperform it. For valence, though, 20 hours of adaptation material was required to catch up with the monolingual 100-hours model.

Of course, one might expect arousal prediction to be quite easy to carry over to another language, while doing this for valence is expected to be a lot harder. However, this is not what we experienced: the prediction of valence (and dominance) was harder than that of arousal {\it even for the within-language model}. In our opinion this might be due to the fact that mostly English data was used to pre-train the WavLM Large model, although this hypothesis is really hard to verify.

Regarding the limitations of our study, we performed the experiments only for two, although very different languages. 
Another limitation can be that we employed only one DNN architecture, and investigated only the emotional attributes. In the near future we plan to extend our tests with categorical emotion recognition experiments, and with incorporating other languages as well, provided that we can obtain large enough corpora which have annotations consistent to the MSP-Podcast and BIIC-Podcast corpora.

\newpage

\section{Acknowledgments}

\ifcameraready
 
This research was supported in part by the Ministry of Culture and Innovation of Hungary from the National Research, Development and Innovation Fund, financed under the a 2024-1.2.3-HU-RIZONT funding scheme (project no. 2024-1.2.3-HU-RIZONT-2024-00017).
\else
placeholder placeholder placeholder placeholder placeholder placeholder placeholder placeholder placeholder placeholder placeholder placeholder placeholder placeholder placeholder placeholder placeholder placeholder placeholder placeholder placeholder placeholder placeholder placeholder placeholder placeholder placeholder placeholder placeholder placeholder placeholder placeholder placeholder placeholder placeholder placeholder placeholder 
\fi

\section{Generative AI Use Disclosure}

Generative AI (Chat GPT and Gemini) was used to correct grammatical and stylistic errors in this paper. 
We did not use any generative AI tools in the design of the study, conducting the experiment, or interpreting the results.

\bibliographystyle{IEEEtran}
\bibliography{mybib}

\end{document}